\input harvmac
\noblackbox

\input epsf

\newcount\figno
\figno=0
\def\fig#1#2#3{
\par\begingroup\parindent=0pt\leftskip=1cm\rightskip=1cm\parindent=0pt
\baselineskip=11pt
\global\advance\figno by 1
\midinsert
\epsfxsize=#3
\centerline{\epsfbox{#2}}
\vskip 12pt
{\bf Fig.\ \the\figno: } #1\par
\endinsert\endgroup\par
}
\def\figlabel#1{\xdef#1{\the\figno}}
\def\encadremath#1{\vbox{\hrule\hbox{\vrule\kern8pt\vbox{\kern8pt
\hbox{$\displaystyle #1$}\kern8pt}
\kern8pt\vrule}\hrule}}

\def\apm{{\alpha^{\prime}}}

\font\cmss=cmss10
\font\cmsss=cmss10 at 7pt
\def\rlx{\relax\leavevmode}
\def\inbar{\vrule height1.5ex width.4pt depth0pt}
\def\IN{\relax{\rm I\kern-.18em N}}
\def\IP{\relax{\rm I\kern-.18em P}}
\def\ZZ{\rlx\leavevmode\ifmmode\mathchoice{\hbox{\cmss Z\kern-.4em Z}}
 {\hbox{\cmss Z\kern-.4em Z}}{\lower.9pt\hbox{\cmsss Z\kern-.36em Z}}
 {\lower1.2pt\hbox{\cmsss Z\kern-.36em Z}}\else{\cmss Z\kern-.4em
 Z}\fi}
\def\IZ{\relax\ifmmode\mathchoice
{\hbox{\cmss Z\kern-.4em Z}}{\hbox{\cmss Z\kern-.4em Z}}
{\lower.9pt\hbox{\cmsss Z\kern-.4em Z}}
{\lower1.2pt\hbox{\cmsss Z\kern-.4em Z}}\else{\cmss Z\kern-.4em
Z}\fi}
\def\IZ{\relax\ifmmode\mathchoice
{\hbox{\cmss Z\kern-.4em Z}}{\hbox{\cmss Z\kern-.4em Z}}
{\lower.9pt\hbox{\cmsss Z\kern-.4em Z}}
{\lower1.2pt\hbox{\cmsss Z\kern-.4em Z}}\else{\cmss Z\kern-.4em
Z}\fi}

\def\narrowplus{\kern -.04truein + \kern -.03truein}
\def\narrowminus{- \kern -.04truein}
\def\narrowminussub{\kern -.02truein - \kern -.01truein}

\def\t{{\theta}}

\def\s{{\sigma}}
\def\frac#1#2{{#1\over #2}}

\def\IZ{\relax\ifmmode\mathchoice
{\hbox{\cmss Z\kern-.4em Z}}{\hbox{\cmss Z\kern-.4em Z}}
{\lower.9pt\hbox{\cmsss Z\kern-.4em Z}}
{\lower1.2pt\hbox{\cmsss Z\kern-.4em Z}}\else{\cmss Z\kern-.4em
Z}\fi}
\def\IC{{\relax\,\hbox{$\inbar\kern-.3em{\rm C}$}}}
\def\p{\partial}
\font\cmss=cmss10 \font\cmsss=cmss10 at 7pt
\def\IR{\relax{\rm I\kern-.18em R}}
\def\ra{\rangle}
\def\la{\langle}

\def\IZ{\relax\ifmmode\mathchoice
{\hbox{\cmss Z\kern-.4em Z}}{\hbox{\cmss Z\kern-.4em Z}}
{\lower.9pt\hbox{\cmsss Z\kern-.4em Z}}
{\lower1.2pt\hbox{\cmsss Z\kern-.4em Z}}\else{\cmss Z\kern-.4em
Z}\fi}
\def\IC{{\relax\,\hbox{$\inbar\kern-.3em{\rm C}$}}}

\def\D{{\Delta}}
\def\m{{\mu}}

\def\d{{\delta}}

\def\G{{\Gamma}}

\def\t{{\theta}}

\def\P{{\Phi}}

\def\t{{\tau}}
\def\r{{\rho}}
\def\th{{\theta}}

\def\s{{\sigma}}

\def\frac#1#2{{#1\over #2}}

\def\p{\partial}

\def\apm{\alpha^{\prime}}

\lref\th{G.~'tHooft, ``A Planar Diagram Theory for Strong Interactions,'' Nucl.\ Phys.\
{\bf 72}, 461, (1974).}
\lref\gv{
R.~Gopakumar and C.~Vafa,
``On the gauge theory/geometry correspondence,''
Adv.\ Theor.\ Math.\ Phys.\  {\bf 3}, 1415 (1999)
[arXiv:hep-th/9811131].}
\lref\ov{
H.~Ooguri and C.~Vafa,
``Worldsheet derivation of a large N duality,''
Nucl.\ Phys.\ B {\bf 641}, 3 (2002)
[arXiv:hep-th/0205297].}
\lref\bov{N.~Berkovits, H.~Ooguri and C.~Vafa,
``On the worldsheet derivation of large N dualities for the superstring,''
arXiv:hep-th/0310118.}
\lref\malda{
J.~M.~Maldacena,
``The large $N$ limit of superconformal field theories and supergravity,''
Adv.\ Theor.\ Math.\ Phys.\  {\bf 2}, 231 (1998)
[Int.\ J.\ Theor.\ Phys.\  {\bf 38}, 1113 (1999)]
[arXiv:hep-th/9711200].}
\lref\ff{R.~Gopakumar,
``From free fields to AdS,''
arXiv:hep-th/0308184.}
\lref\divec{P.~Di Vecchia, L.~Magnea, A.~Lerda, R.~Russo and R.~Marotta,
``String techniques for the calculation of renormalization constants in field theory,''
Nucl.\ Phys.\ B {\bf 469}, 235 (1996)
[arXiv:hep-th/9601143].}
\lref\fmr{A.~Frizzo, L.~Magnea and R.~Russo,
``Systematics of one-loop Yang-Mills diagrams from bosonic string  amplitudes,''
Nucl.\ Phys.\ B {\bf 604}, 92 (2001)
[arXiv:hep-ph/0012129].}
\lref\berkos{
Z.~Bern and D.~A.~Kosower,
``Efficient Calculation Of One Loop QCD Amplitudes,''
Phys.\ Rev.\ Lett.\  {\bf 66}, 1669 (1991).}
\lref\bdh{
L.~Brink, P.~Di Vecchia and P.~S.~Howe,
``A Lagrangian Formulation Of The Classical And Quantum Dynamics Of Spinning Particles,''
Nucl.\ Phys.\ B {\bf 118}, 76 (1977).}
\lref\ssen{M.~Tuite and S.~Sen,
``A String Motivated Approach to the Relativistic Point Particle,''
arXiv:hep-th/0308099.}
\lref\strass{M.~J.~Strassler,
``Field theory without Feynman diagrams: One loop effective actions,''
Nucl.\ Phys.\ B {\bf 385}, 145 (1992)
[arXiv:hep-ph/9205205].}
\lref\schb{C.~Schubert,
``Perturbative quantum field theory in the string-inspired formalism,''
Phys.\ Rept.\  {\bf 355}, 73 (2001)
[arXiv:hep-th/0101036].}
\lref\mikh{A.~Mikhailov,
``Notes on higher spin symmetries,''
arXiv:hep-th/0201019.}
\lref\cnss{G.~Chalmers, H.~Nastase, K.~Schalm and R.~Siebelink,
``R-current correlators in N = 4 super Yang-Mills theory from anti-de  Sitter supergravity,''
Nucl.\ Phys.\ B {\bf 540}, 247 (1999)
[arXiv:hep-th/9805105].}
\lref\guil{E. A. Guillemin, ``Introductory Circuit Theory'', (John Wiley and Sons, 1953).}
\lref\BD{J. D. Bjorken and S. D. Drell, ``Relativistic Quantum Fields'', (McGraw Hill, 1965).}
\lref\iz{C.~Itzykson and J-B.~Zuber, ``Quantum Field Theory,'' (Mc Graw Hill, (1980).}
\lref\lam{C.~S.~Lam,
``Multiloop string - like formulas for QED,''
Phys.\ Rev.\ D {\bf 48}, 873 (1993)
[arXiv:hep-ph/9212296].}
\lref\lama{C.~S.~Lam,
``Spinor helicity technique and string reorganization for multiloop
diagrams,''
Can.\ J.\ Phys.\  {\bf 72}, 415 (1994)
[arXiv:hep-ph/9308289].}
\lref\petkou{A.~Petkou,
``Conserved currents, consistency relations, and operator product  expansions
in the conformally invariant O(N) vector model,''
Annals Phys.\  {\bf 249}, 180 (1996)
[arXiv:hep-th/9410093].}
\lref\liu{H.~Liu,
``Scattering in anti-de Sitter space and operator product expansion,''
Phys.\ Rev.\ D {\bf 60}, 106005 (1999)
[arXiv:hep-th/9811152].}
\lref\dmmr{E.~D'Hoker, S.~D.~Mathur, A.~Matusis and L.~Rastelli,
``The operator product expansion of N = 4 SYM and the 4-point functions  of supergravity,''
Nucl.\ Phys.\ B {\bf 589}, 38 (2000)
[arXiv:hep-th/9911222].}
\lref\dfmmr{E.~D'Hoker, D.~Z.~Freedman, S.~D.~Mathur, A.~Matusis and L.~Rastelli,
``Graviton exchange and complete 4-point functions in the AdS/CFT  correspondence,''
Nucl.\ Phys.\ B {\bf 562}, 353 (1999)
[arXiv:hep-th/9903196].}
\lref\fmmr{D.~Z.~Freedman, S.~D.~Mathur, A.~Matusis and L.~Rastelli,
``Correlation functions in the CFT($d$)/AdS($d+1$) correspondence,''
Nucl.\ Phys.\ B {\bf 546}, 96 (1999)
[arXiv:hep-th/9804058].}
\lref\sunda{B.~Sundborg,
``The Hagedorn transition, deconfinement and N = 4 SYM theory,''
Nucl.\ Phys.\ B {\bf 573}, 349 (2000)
[arXiv:hep-th/9908001].}
\lref\sund{P.~Haggi-Mani and B.~Sundborg,
``Free large N supersymmetric Yang-Mills theory as a string theory,''
JHEP {\bf 0004}, 031 (2000)
[arXiv:hep-th/0002189];}
\lref\sundb{B.~Sundborg,
``Stringy gravity, interacting tensionless strings and massless higher  spins,''
Nucl.\ Phys.\ Proc.\ Suppl.\  {\bf 102}, 113 (2001)
[arXiv:hep-th/0103247].}
\lref\witt{E. Witten, Talk at the Schwarzfest,
http://theory.caltech.edu/jhs60/witten.}
\lref\tset{A.~A.~Tseytlin,
``On limits of superstring in AdS(5) x S**5,''
Theor.\ Math.\ Phys.\  {\bf 133}, 1376 (2002)
[Teor.\ Mat.\ Fiz.\  {\bf 133}, 69 (2002)]
[arXiv:hep-th/0201112].}
\lref\mz{J.~A.~Minahan and K.~Zarembo,
``The Bethe-ansatz for N = 4 super Yang-Mills,''
JHEP {\bf 0303}, 013 (2003)
[arXiv:hep-th/0212208].}
\lref\dmw{A.~Dhar, G.~Mandal and S.~R.~Wadia,
``String bits in small radius AdS and weakly coupled N = 4 super  Yang-Mills theory. I,''
arXiv:hep-th/0304062.}
\lref\dnw{L.~Dolan, C.~R.~Nappi and E.~Witten,
``A relation between approaches to integrability in superconformal Yang-Mills theory,''
arXiv:hep-th/0308089.}
\lref\polch{J.~Polchinski, Seminar at SW workshop on String Theory (Feb. 2003), unpublished.}
\lref\msw{G.~Mandal, N.~V.~Suryanarayana and S.~R.~Wadia,
``Aspects of semiclassical strings in AdS(5),''
Phys.\ Lett.\ B {\bf 543}, 81 (2002)
[arXiv:hep-th/0206103].}
\lref\bnp{I.~Bena, J.~Polchinski and R.~Roiban,
``Hidden symmetries of the AdS(5) x S**5 superstring,''
arXiv:hep-th/0305116.}
\lref\vall{B.~C.~Vallilo,
``Flat currents in the classical AdS(5) x S**5 pure spinor superstring,''
arXiv:hep-th/0307018.}
\lref\klepol{I.~R.~Klebanov and A.~M.~Polyakov,
``AdS dual of the critical O(N) vector model,''
Phys.\ Lett.\ B {\bf 550}, 213 (2002)
[arXiv:hep-th/0210114].}
\lref\peta{A.~C.~Petkou,
``Evaluating the AdS dual of the critical O(N) vector model,''
JHEP {\bf 0303}, 049 (2003)
[arXiv:hep-th/0302063].}
\lref\leigh{R.~G.~Leigh and A.~C.~Petkou,
``Holography of the N = 1 higher-spin theory on AdS(4),''
JHEP {\bf 0306}, 011 (2003)
[arXiv:hep-th/0304217].}
\lref\sezsuna{E.~Sezgin and P.~Sundell,
``Holography in 4D (super) higher spin theories and a test via cubic  scalar couplings,''
arXiv:hep-th/0305040.}
\lref\por{L.~Girardello, M.~Porrati and A.~Zaffaroni,
``3-D interacting CFTs and generalized Higgs phenomenon in higher spin  theories on AdS,''
Phys.\ Lett.\ B {\bf 561}, 289 (2003)
[arXiv:hep-th/0212181].}
\lref\ruhl{T.~Leonhardt, A.~Meziane and W.~Ruhl,
``On the proposed AdS dual of the critical O(N) sigma model for any  dimension $2 < d < 4$,''
Phys.\ Lett.\ B {\bf 555}, 271 (2003)
[arXiv:hep-th/0211092].}
\lref\rajan{F.~Kristiansson and P.~Rajan,
``Scalar field corrections to AdS(4) gravity from higher spin gauge  theory,''
JHEP {\bf 0304}, 009 (2003)
[arXiv:hep-th/0303202].}
\lref\sumit{S.~R.~Das and A.~Jevicki,
``Large-N collective fields and holography,''
Phys.\ Rev.\ D {\bf 68}, 044011 (2003)
[arXiv:hep-th/0304093].}
\lref\nemani{N.~V.~Suryanarayana,
JHEP {\bf 0306}, 036 (2003)
[arXiv:hep-th/0304208].}
\lref\pol{A.~M.~Polyakov,
``Gauge fields and space-time,''
Int.\ J.\ Mod.\ Phys.\ A {\bf 17S1}, 119 (2002)
[arXiv:hep-th/0110196].}
\lref\polb{A.~M.~Polyakov,
``String theory and quark confinement,''
Nucl.\ Phys.\ Proc.\ Suppl.\  {\bf 68}, 1 (1998)
[arXiv:hep-th/9711002].}
\lref\polc{A.~M.~Polyakov,
``The wall of the cave,''
Int.\ J.\ Mod.\ Phys.\ A {\bf 14}, 645 (1999)
[arXiv:hep-th/9809057].}
\lref\polbk{A.~M.~Polyakov,
``Gauge fields and Strings,'' (Harwood Academic Publishers, 1987).}
\lref\bmn{D.~Berenstein, J.~M.~Maldacena and H.~Nastase,
``Strings in flat space and pp waves from N = 4 super Yang Mills,''
JHEP {\bf 0204}, 013 (2002)
[arXiv:hep-th/0202021].}
\lref\gkp{S.~S.~Gubser, I.~R.~Klebanov and A.~M.~Polyakov,
``Gauge theory correlators from non-critical string theory,''
Phys.\ Lett.\ B {\bf 428}, 105 (1998)
[arXiv:hep-th/9802109].}
\lref\witads{E.~Witten,
``Anti-de Sitter space and holography,''
Adv.\ Theor.\ Math.\ Phys.\  {\bf 2}, 253 (1998)
[arXiv:hep-th/9802150].}
\lref\wittwist{E.~Witten,
``Perturbative gauge theory as a string theory in twistor space,''
arXiv:hep-th/0312171.}
\lref\berktwist{N.~Berkovits,
``An Alternative String Theory in Twistor Space for N=4 Super-Yang-Mills,''
arXiv:hep-th/0402045.}
\lref\rsv{R.~Roiban, M.~Spradlin and A.~Volovich,
``A googly amplitude from the B-model in twistor space,''
arXiv:hep-th/0402016.}
\lref\sezsun{E.~Sezgin and P.~Sundell,
``Doubletons and 5D higher spin gauge theory,''
JHEP {\bf 0109}, 036 (2001)
[arXiv:hep-th/0105001]; ``Massless higher spins and holography,''
Nucl.\ Phys.\ B {\bf 644}, 303 (2002)
[Erratum-ibid.\ B {\bf 660}, 403 (2003)]
[arXiv:hep-th/0205131].}
\lref\vas{
M.~A.~Vasiliev,
``Conformal higher spin symmetries of 4D massless supermultiplets and  osp(L,2M)
invariant equations in generalized (super)space,''
Phys.\ Rev.\ D {\bf 66}, 066006 (2002)
[arXiv:hep-th/0106149].}
\lref\vasrev{M.~A.~Vasiliev,
``Higher spin gauge theories: Star-product and AdS space,''
arXiv:hep-th/9910096.}
\lref\bv{A.~O.~Barvinsky and G.~A.~Vilkovisky,
``Beyond The Schwinger-Dewitt Technique: Converting Loops Into Trees And In-In Currents,''
Nucl.\ Phys.\ B {\bf 282}, 163 (1987).}
\lref\birdav{N.~D.~Birrell and P.~C.~W.~Davies, ``Quantum fields in Curved Space,'' Cambridge University
Press (1984).}
\lref\hsken{M.~Henningson and K.~Skenderis,
``The holographic Weyl anomaly,''
JHEP {\bf 9807}, 023 (1998)
[arXiv:hep-th/9806087].}
\lref\nojod{S.~Nojiri and S.~D.~Odintsov,
``Conformal anomaly for dilaton coupled theories from AdS/CFT  correspondence,''
Phys.\ Lett.\ B {\bf 444}, 92 (1998)
[arXiv:hep-th/9810008].}
\lref\odnoj{S.~Nojiri, S.~D.~Odintsov and S.~Ogushi,
``Finite action in d5 gauged supergravity and dilatonic conformal anomaly  for dual quantum field theory,''
Phys.\ Rev.\ D {\bf 62}, 124002 (2000)
[arXiv:hep-th/0001122].}
\lref\dss{S.~de Haro, S.~N.~Solodukhin and K.~Skenderis,
``Holographic reconstruction of spacetime and renormalization in the  AdS/CFT correspondence,''
Commun.\ Math.\ Phys.\  {\bf 217}, 595 (2001)
[arXiv:hep-th/0002230].}
\lref\feffgr{C.~Fefferman, C.~R.~Graham, ``Conformal Invariants'' in ``{\it Elie Cartan et les
Mathematiques d'Aujourd'hui},'' (Asterisque 1985), 95.}
\lref\glebfrpet{G.~Arutyunov, S.~Frolov and A.~C.~Petkou,
``Operator product expansion of the lowest weight CPOs in N = 4  SYM(4) at strong coupling,''
Nucl.\ Phys.\ B {\bf 586}, 547 (2000)
[Erratum-ibid.\ B {\bf 609}, 539 (2001)]
[arXiv:hep-th/0005182];
``Perturbative and instanton corrections to the OPE of CPOs in N = 4  SYM(4),''
Nucl.\ Phys.\ B {\bf 602}, 238 (2001)
[Erratum-ibid.\ B {\bf 609}, 540 (2001)]
[arXiv:hep-th/0010137].}
\lref\arutfrol{G.~Arutyunov and S.~Frolov,
``Three-point Green function of the stress-energy tensor in the AdS/CFT  correspondence,''
Phys.\ Rev.\ D {\bf 60}, 026004 (1999)
[arXiv:hep-th/9901121].}
\lref\east{M.~G.~Eastwood,
``Higher symmetries of the Laplacian,''
arXiv:hep-th/0206233.}
\lref\gm{D.~J.~Gross and P.~F.~Mende,
``String Theory Beyond The Planck Scale,''
Nucl.\ Phys.\ B {\bf 303}, 407 (1988).}
\lref\gross{D.~J.~Gross,
``High-Energy Symmetries Of String Theory,''
Phys.\ Rev.\ Lett.\  {\bf 60}, 1229 (1988).}
\lref\bianchi{M.~Bianchi, J.~F.~Morales and H.~Samtleben,
``On stringy AdS(5) x S**5 and higher spin holography,''
JHEP {\bf 0307}, 062 (2003).}
\lref\bbms{N.~Beisert, M.~Bianchi, J.~F.~Morales and H.~Samtleben,
``On the spectrum of AdS/CFT beyond supergravity,''
arXiv:hep-th/0310292.}
\lref\anselmi{D.~Anselmi,
``The N = 4 quantum conformal algebra,''
Nucl.\ Phys.\ B {\bf 541}, 369 (1999)
[arXiv:hep-th/9809192].}
\lref\lind{U.~Lindstrom and M.~Zabzine,
``Tensionless strings, WZW models at critical level and massless higher  spin fields,''
arXiv:hep-th/0305098.}
\lref\witcube{E.~Witten,
``Noncommutative Geometry And String Field Theory,''
Nucl.\ Phys.\ B {\bf 268}, 253 (1986).}
\lref\gmw{S.~B.~Giddings, E.~J.~Martinec and E.~Witten,
``Modular Invariance In String Field Theory,''
Phys.\ Lett.\ B {\bf 176}, 362 (1986).}
\lref\zwie{B.~Zwiebach,
``A Proof That Witten's Open String Theory Gives A Single Cover Of Moduli
Space,''
Commun.\ Math.\ Phys.\  {\bf 142}, 193 (1991).}
\lref\penn{R.~Penner, ``Perturbative Series and the Moduli Space of
Riemann Surfaces,'' J. \ Diff.\ Geom.{\bf 27}, 35 (1988).}
\lref\sunil{S.~Mukhi,
``Topological matrix models, Liouville matrix model and c = 1 string theory,''
arXiv:hep-th/0310287.}
\lref\ashoke{A.~Sen, ``Open-closed duality at tree level,''
arXiv:hep-th/0306137.} 
\lref\gir{D.~Gaiotto, N.~Itzhaki and
L.~Rastelli, ``Closed strings as imaginary D-branes,''
arXiv:hep-th/0304192.} 
\lref\grsz{D.~Gaiotto, L.~Rastelli, A.~Sen and B.~Zwiebach,
``Ghost structure and closed strings in vacuum string field theory,''
Adv.\ Theor.\ Math.\ Phys.\  {\bf 6}, 403 (2003)
[arXiv:hep-th/0111129].}
\lref\gr{D.~Gaiotto and L.~Rastelli,
 ``A paradigm of open/closed duality: Liouville D-branes and the Kontsevich
model,''
arXiv:hep-th/0312196.}
\lref\satchi{S.~R.~Das, S.~Naik and S.~R.~Wadia, ``Quantization Of
The Liouville Mode And String Theory,'' Mod.\ Phys.\ Lett.\ A {\bf
4}, 1033 (1989).} 
\lref\spenta{A.~Dhar, T.~Jayaraman, K.~S.~Narain
and S.~R.~Wadia, ``The Role Of Quantized Two-Dimensional Gravity
In String Theory,'' Mod.\ Phys.\ Lett.\ A {\bf 5}, 863 (1990).}
\lref\karch{A.~Clark, A.~Karch, P.~Kovtun and D.~Yamada,
``Construction of bosonic string theory on infinitely curved
anti-de  Sitter space,'' arXiv:hep-th/0304107; A.~Karch,
``Lightcone quantization of string theory duals of free field
theories,'' arXiv:hep-th/0212041.} 
\lref\lmrs{S.~M.~Lee,
S.~Minwalla, M.~Rangamani and N.~Seiberg, ``Three-point functions
of chiral operators in D = 4, N = 4 SYM at  large N,'' Adv.\
Theor.\ Math.\ Phys.\  {\bf 2}, 697 (1998)
[arXiv:hep-th/9806074].} 
\lref\hsw{P.~S.~Howe, E.~Sokatchev and P.~C.~West,
``3-point functions in N = 4 Yang-Mills,''
Phys.\ Lett.\ B {\bf 444}, 341 (1998)
[arXiv:hep-th/9808162].}
\lref\dfs{E.~D'Hoker, D.~Z.~Freedman and W.~Skiba,
``Field theory tests for correlators in the AdS/CFT correspondence,''
Phys.\ Rev.\ D {\bf 59}, 045008 (1999)
[arXiv:hep-th/9807098].}
\lref\mettse{R.~R.~Metsaev and
A.~A.~Tseytlin, ``Type IIB superstring action in AdS(5) x S(5)
background,'' Nucl.\ Phys.\ B {\bf 533}, 109 (1998)
[arXiv:hep-th/9805028].} 
\lref\berk{N.~Berkovits, ``Super-Poincare
covariant quantization of the superstring,'' JHEP {\bf 0004}, 018
(2000) [arXiv:hep-th/0001035].} 
\lref\polchi{J.~Polchinski, Nucl.\
Phys.\ B {\bf 331}, 123, (1989).}
\lref\ymtherm{O.~Aharony, J.~Marsano, S.~Minwalla, K.~Papadodimas 
and M.~Van Raamsdonk,
``The Hagedorn / deconfinement phase transition in weakly coupled large N
gauge theories,''
arXiv:hep-th/0310285.}
\lref\thorn{K.~Bardakci and C.~B.~Thorn,
Nucl.\ Phys.\ B {\bf 626}, 287 (2002)
[arXiv:hep-th/0110301]; C.~B.~Thorn,
Nucl.\ Phys.\ B {\bf 637}, 272 (2002)
[Erratum-ibid.\ B {\bf 648}, 457 (2003)]
[arXiv:hep-th/0203167]; K.~Bardakci and C.~B.~Thorn,
Nucl.\ Phys.\ B {\bf 652}, 196 (2003)
[arXiv:hep-th/0206205]; S.~Gudmundsson, C.~B.~Thorn and T.~A.~Tran,
Nucl.\ Phys.\ B {\bf 649}, 3 (2003)
[arXiv:hep-th/0209102]; K.~Bardakci and C.~B.~Thorn,
Nucl.\ Phys.\ B {\bf 661}, 235 (2003)
[arXiv:hep-th/0212254]; C.~B.~Thorn and T.~A.~Tran,
Nucl.\ Phys.\ B {\bf 677}, 289 (2004)
[arXiv:hep-th/0307203].}
\lref\kumar{P.~de Medeiros and S.~P.~Kumar,
``Spacetime Virasoro algebra from strings on zero radius AdS(3),''
JHEP {\bf 0312}, 043 (2003)
[arXiv:hep-th/0310040].}
\lref\tse{A.~A.~Tseytlin,
``On semiclassical approximation and spinning string vertex operators in
AdS(5) x S**5,''
Nucl.\ Phys.\ B {\bf 664}, 247 (2003)
[arXiv:hep-th/0304139].}

\Title
{\vbox{\baselineskip12pt
\hbox{hep-th/0402063}}}
{\vbox{\centerline{From Free Fields to $AdS$ -- II}}}

\centerline{Rajesh Gopakumar\foot{gopakumr@mri.ernet.in}}

\centerline{\sl $^a$Harish-Chandra Research Institute, Chhatnag Rd.,}
\centerline{\sl Jhusi, Allahabad, India 211019.}
\centerline{\sl $^b$ Dept. of Physics, I.I.T. Kanpur,}
\centerline{\sl Kanpur, India 208016.}
\medskip

\vskip 0.8cm

\centerline{\bf Abstract}
\medskip
\noindent

We continue with the program of hep-th/0308184 to implement
open-closed string duality on free gauge field 
theory (in the large $N$ limit). In this paper we consider
correlators such as $\la \prod_{i=1}^n \Tr\Phi^{J_i}(x_i)\ra$. The
Schwinger parametrisation of this $n$-point function exhibits a
partial gluing up into a set of basic skeleton graphs. We argue
that the moduli space of the planar skeleton graphs is exactly the same
as the moduli space of genus zero Riemann surfaces with $n$ holes.
In other words, we can explicitly rewrite the $n$-point (planar) free field
correlator as an integral over the moduli space of a sphere with
$n$ holes. A preliminary study of the integrand also indicates 
compatibility with a string theory on $AdS$. 
The details of our argument are
quite insensitive to the specific form of the operators and
generalise to diagrams of higher genus as well. We take this as
evidence of the field theory's ability to reorganise itself into a
string theory.

\vskip 0.5cm
\Date{February 2004}
\listtoc
\writetoc

\newsec{Introduction}

How exactly does a quantum field theory (in the large $N$ limit)
reassemble itself into a closed string theory? This question lies at
the heart of the gauge theory/geometry correspondence. Answering
it in its generality is likely to give us valuable clues regarding the
string dual to $QCD$, for instance.

What we have learnt in the years
since Maldacena's breakthrough is that the answer to this
question is tied up with open-closed string duality. The gauge theory
arising in an open string description is related by worldsheet
duality to a closed string description. The holes of the open string
worldsheet get glued up, getting replaced by closed string insertions.

In the case of topological string dualities
it was possible \gv\ov\ to make this
intuition precise  using a linear sigma model description of the
worldsheet (the argument for the corresponding F-terms in the superstring
was made in \bov ).
 Recently, a very nice illustration of open-closed string
duality was given, again in a topological context, for the Kontsevich
matrix model \gr . Here again,
one could  concretely see the process of holes
closing up and being replaced by closed string insertions.

Nevertheless, the original AdS/CFT conjecture \malda\gkp\witads\
has not yet been understood in such terms.\foot{The  recent
proposals \wittwist\berktwist (see also \rsv ) 
that ${\cal N}=4$ Yang-Mills arises as the
target space theory of a topological sigma model might, perhaps,
enable one to view it in a manner close to the other topological
examples.} In \ff\ we embarked on an effort to implement
open-closed string duality in the free field limit of the ${\cal
N}=4$ Super Yang-Mills theory. The tractability of the limit, from
the  field theory point of view, makes it a natural 
starting point.\foot{See 
\sund\sunda\sundb\pol\witt\mikh\tset\bianchi\ymtherm\bbms
for various investigations of the free/weakly coupled
theory with a view to understanding its stringy dual. Another approach 
starting from light cone field theory is that of Thorn and 
collaborators \thorn\ as well as that of Karch and collaborators \karch . 
There is also a lot of literature on the connection
between weakly coupled ${\cal N}=4$ Yang-Mills theory and 
integrable spin chains, 
since the work of \mz .} The
strategy in \ff\ was to consider a worldline representation
(Schwinger parametrisation) of the free field correlators. This
was motivated by the fact that these representations can be viewed
as being directly inherited from the relevant open string theory in the
$\apm\rightarrow\infty$ limit. A nice feature of this
representation is its correspondence with electrical networks.
This correspondence suggested that carrying out the integration
over the internal loop momenta (eliminating internal currents)
should yield an equivalent network, now with a tree-like
structure. In other words, the holes would have been closed up.
The idea was then, through a change of variables on the Schwinger
moduli space, to exhibit the integral as that of a closed string
tree amplitude on $AdS$.

In \ff\ we restricted ourselves to bilinear operators (such as
$\Tr\Phi^2$). The $n$ point function of these operators is given
by a one loop diagram. For the case of two and three point
functions the equivalent tree diagrams are the expected ones. A
simple change of variables on the Schwinger parameters converted
the integral to a tree amplitude in AdS. We further gave arguments
for the four point function that the resulting tree structure is
again in line with expectations, though a detailed check was not
carried out.

In the present paper we will consider a much more general class of
operators and their correlators such as\foot{As in \ff\ we will be considering 
a $U(N)$ Euclidean gauge field theory. We will again be dropping factors which
are ``inessential'', in all equations.} 
\eqn\ncorr{
G^{\{J_i\}}(x_1, x_2,\ldots x_n)=
\la \prod_{i=1}^n \Tr\Phi^{J_i}(x_i)\ra_{conn}. }
All possible free wick contractions lead to a large class of diagrams
contributing to such a correlator, even if one restricts to planar graphs.
These diagrams have $n$ vertices
with $J_i$ legs coming out of the $i$'th vertex.
We will argue, from the Schwinger parametrised expressions
for such diagrams, that they exhibit a (partial) gluing up into
a skeleton diagram (with $n$ vertices)
which captures the basic connectivity of the original
graph. This is illustrated in Fig.1.

\fig{Gluing up of a planar six point function into a skeleton graph.}
{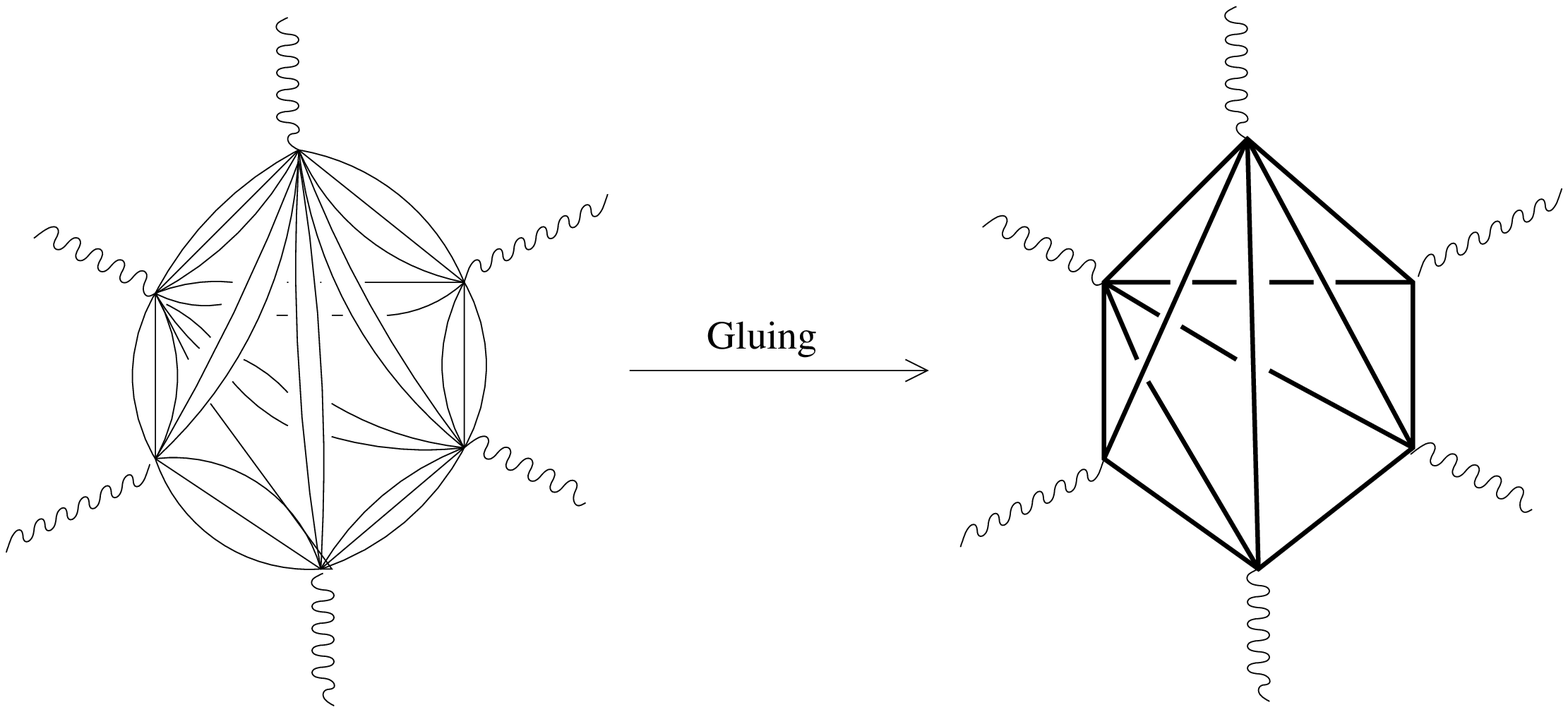}{6.0truein}

This gluing can be intuitively understood from the electrical
analogy since it essentially involves replacing the various parallel
resistors (Schwinger parameters), between a pair of vertices, with
a single effective resistor. Therefore, any particular
contribution to the $n$-point function \ncorr\ can be expressed as
an integral over a reduced Schwinger parameter space, namely that
of the corresponding skeleton graph. The information about the
$J_i$ is captured through a specific dependence in the integrand.

However, planar graphs with different connectivities give rise to
different skeleton diagrams. All these different skeleton diagram
contributions need to be summed over to obtain the complete
answer for \ncorr . We will argue that this space of skeleton
graphs is in one-to-one correspondence with the familiar cell
decomposition of the moduli space ${\cal M}_{0,n}$, of a sphere
with $n$ holes. This basically follows from considering the graphs
which are dual (in the graph theory sense) to the skeleton
diagrams.

It is important to stress that this moduli space is {\it distinct}
from that of the string diagrams underlying the original field
theory Feynman diagrams. As evident from the contributions to
\ncorr\ shown in Fig.1, these have a large number of loops (the
number depending on $J_i$). In fact, even the skeleton graphs
themselves have (generically) $2(n-2)$ faces. Whereas, the moduli
space we are associating with all $n$-point correlators such as in
\ncorr , is that of a sphere with exactly $n$ holes. Moreover,
field theory correlators typically get their contribution from
corners of string moduli space whereas here it is the full
moduli space ${\cal M}_{0,n}$ which contributes. Thus this stringy 
representation of field theory
is different from that studied by Bern, Kosower \berkos\
and others.

In fact, the emergence of the moduli space of a sphere with $n$
holes is natural from the point of view of the gauge
theory/geometry correspondence. The scenario one expects is that
the loops of the original field theory planar diagram get glued up
to form a surface and one has instead $n$ closed string
insertions. The $n$ holes that we see here are to be identified
with these closed string insertions. On integrating over the
moduli corresponding to the size of these $n$-holes, the holes
should effectively pinch off giving rise to $n$ external closed
string insertions at punctures. This is indeed the picture that is
realised in the topological string dualities of \gv\ov\gr . The
situation is depicted in figure 2. We will see evidence that this is 
realised in our case, both by looking at the three point 
function in detail, as well as by studying the form of general 
stringy correlators in $AdS$.

\fig{Skeleton graphs $\rightarrow$ sphere with holes $\rightarrow$ sphere with
punctures (as the holes go to $\infty$).}
{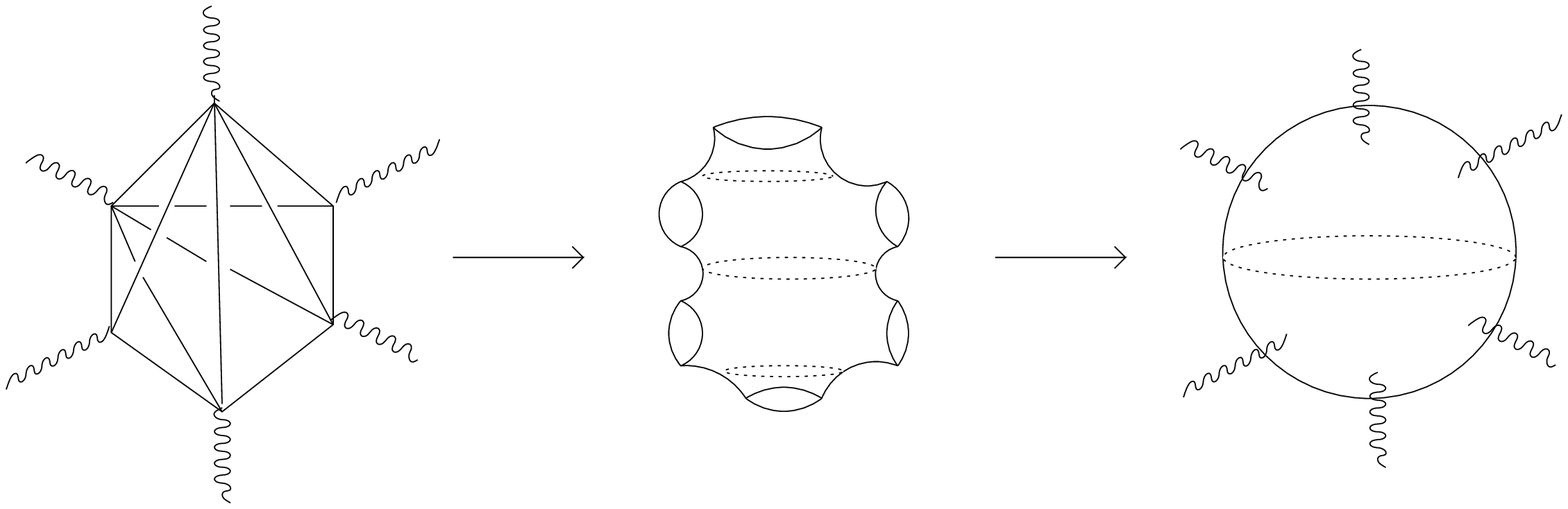}{6.5truein}

It is mainly for the sake of simplicity that we make our arguments 
for the $n$-point correlators of scalars. 
The Schwinger parametrisations for other $n$-point functions in
free field theory is very similar. In particular, the Feynman
graphs get glued up, for exactly the same reasons, into the same
skeleton diagrams. And hence replaying the arguments, we
can conclude that other $n$-point correlators in the free theory
can also be written in terms of an integral over ${\cal M}_{0,n}$.

Moreover, the argument is not
restricted to planar diagrams alone. One can generalise
to diagrams of arbitrary genus which also get glued into skeleton
diagrams. This time one makes a correspondence with the cell decomposition
of the moduli space  ${\cal M}_{g,n}$. Thus in all cases an $n$-point function
leads to a Riemann surface with $n$ holes.
This leads us to believe that
what we are seeing is a signature of the string dual of the free field
theory. In fact, a plus point of this 
procedure is that it is quite likely generalisable to the 
interacting theory as well. 

Having written the field theory expression as an integral over
parameters which cover ${\cal M}_{g,n}$, the main task then
remains to see that the integrand corresponds to that of the
appropriate string theory on $AdS$. In fact, it is tempting to
speculate that the theory on ${\cal M}_{g,n}$, that we are seeing
here, defines {\it a consistent open string theory on a zero size
$AdS$}. The $n$-holes give the contributions of boundary states in
this open string theory. As in the picture of tachyon condensation
(see e.g. \grsz\gir ) or the topological duality of \gr ,\foot{In fact, the 
authors of \gr\ speculate on the existence of such 
an open string theory on $AdS$, in analogy with their example.} integrating
the boundary state over the size modulus of the hole would then
give rise to a description in terms of closed string vertex
operators in $AdS$ inserted at the $n$ punctures. These issues are
under investigation. At present, we will just make a few disparate
remarks in support of the above scenario.

Firstly, as we will see in Sec. 4, in the ``critical'' dimension
$d=4$, the integrand can be written in a particularly nice form as far
as its dependence on the quantum numbers $J_i$ and external momenta $k_i$
go. This form is at least not obviously inconsistent with string theory and
in fact shares many structural features consistent with it, as we will also 
see in Sec. 5.2. An important check is
that the integrand is continuous across the boundaries of the different
components in the cell decomposition of the moduli space.

Secondly, the factorisation of
field theory correlators following from the spacetime OPE should translate into
a factorisation of the amplitudes in the closed string channel. It is
plausible that the integrand in the Schwinger moduli space
should reflect this factorisation and thus provide one of the consistency
checks for it to be a string amplitude. 

Finally, we generalise (in Sec. 5)
the considerations of \ff\ for planar three point functions of bilinears to
that for the more general operators $\Tr\Phi^{J_i}$. We see
in this case that the the three Schwinger
parameters that label ${\cal M}_{0,3}$
(corresponding to the sizes of the three holes) transmute into
parameters for the external legs of $AdS$ propagators. This happens
via the same change  of variables as in \ff . In a
sense, integration over these moduli effectively puts the
insertions at the boundary of $AdS$. 
We will also look at the general form of string correlators 
of scalars in $AdS$ and argue 
that they can very naturally be cast in a form compatible 
with the field theory
expressions obtained in Sec. 4.  
Therefore, together with the emergence of the stringy moduli space 
this gives confidence that
one is implementing the expected picture of
open-closed string duality in this approach.

There are other issues which we do not address directly in this work.
For instance, understanding 
the role of supersymmetry, if any, will have to await 
a more detailed study of the properties of the integrand.
In any case, our firm belief
in the AdS/CFT conjecture tells us that, at least in this case, we
are assured of the free field theory having a closed string dual.
But, in our arguments here, we do not really use any aspect of
${\cal N}=4$ Super Yang-Mills. The procedure thus far is
quite general. Another related issue is that of the
spacetime dimension. The main conclusion of this paper about the
emergence of the moduli space ${\cal M}_{g,n}$ is valid for any
dimension $d$. But the integrand seems to be particularly nice
when $d=4$ (as might be expected of field theories).
A closer examination of the integrand
should reveal more.

The paper is organised as follows. Section 2 displays the
Schwinger parametrisation of correlators such as \ncorr\ and
exhibits their gluing up into skeleton graphs as in Fig.1. Section
3 then makes the correspondence of the parameter space of all
planar skeleton diagrams with the moduli space ${\cal M}_{0,n}$.
It also sketches the generalisations to other correlators as well
as to higher genus. Section 4 makes some general remarks on the integrand
in moduli space as given by field theory. Section 5 studies
the three point function in some detail. It also gives some evidence
for the relation between the field theory integrand and the general
$n$-point stringy correlator on $AdS$.  
Appendix A gives the details 
associated with a change of variables in Sec. 2. 

\newsec{Schwinger Moduli and Skeleton Graphs}

\subsec{A Review of the Parametric Representation}
 
The Schwinger parametric representation of field theory is a well
studied subject. Essentially, one reexpresses the denominator of
all propagators in a Feynman diagram via the identity (appropriate
for Euclidean space correlators) 
\eqn\sch{{1\over
p^2+m^2}=\int_0^{\infty} d\t \exp{\{-\t(p^2+m^2)\}}.} 
This has the
advantage of converting all the momentum integrals into Gaussian
integrals which are then easy to carry out. It is a little
intricate to keep track of the details of the momentum flow. But
the final expressions for an arbitrary Feynman diagram can be
compactly written in graph theoretic terms. For the case of scalar
fields, the expressions can be looked up in field theory textbooks
such as \iz\ (sec.6-2-3). The expressions involving spinors and
gauge fields are more involved. For a recent review containing the
general expressions, see \lam\lama .

Since we will be mostly looking at massless scalar fields, let us 
consider the expression for an arbitrary Feynman diagram contributing 
to the momentum space version of \ncorr .
The result (in $d$ dimensions), of
carrying out the integral over the internal momenta is given in a
(deceptively) compact form\foot{Here we are suppressing the overall momentum 
conserving delta function.}
\eqn\scal{G(k_1, k_2 \ldots k_n)=
\int_0^{\infty}{\prod_{r}d\t_r\over \D (\t)^{d\over
2}}\exp{\{-P(\t, k)\}}.}
Here the product over $r$ goes over all the internal lines in the
graph -- there being one Schwinger parameter for each such line.
The measure factor $\D (\t)$ and the Gaussian exponent $P(\t, k)$
are given by (see for e.g. \iz\lam )
\eqn\meas{\D(\t)=\sum_{T_1}(\prod^{l}\t).} \eqn\gauss{P(\t,
k)=\D(\t)^{-1}\sum_{T_2}(\prod^{l+1}\t)(\sum k)^2.} Here we are
following the notation of \lam : The sum is over various 1-trees
and 2-trees obtained from the original loop diagram. A 1-tree is
obtained by cutting $l$ lines of a diagram with $l$ loops so as to
make a connected tree. While a 2-tree is obtained by cutting $l+1$
lines of the loop so as to form two disjoint trees. \meas\
indicates a sum over the set $T_1$ of all 1-trees, with the
product over the $l$ Schwinger parameters of all the cut lines.
The sum over $T_2$ in \gauss\ similarly indicates a sum over the
set of all two trees, where the product is over the $\t$'s of the
$l+1$ cut lines. And $(\sum k)$ is understood to be the sum over
all those external momenta $k_i$ which flow into (either) one of
the two trees. (Note that because of overall momentum
conservation, it does not matter which set of external momenta one
chooses.)

A simple illustration of these expressions is for the one loop diagram
with $n$ insertions. There are $n$ Schwinger parameters for each of the
$n$ arc segments of this loop. Cutting any of them leads to a 1-tree.
Therefore
$$\D(\t)_{(l=1)}=\sum_{i=1}^n \t_i.$$
Cutting any two distinct ones leads to two disjoint trees and
$$P(\t,k)_{(l=1)}=\D(\t)_{(l=1)}^{-1}\sum_{i<j}\t_i\t_j(k_{i+1}+\ldots k_j)^2,$$
where $\t_i$ is the parameter for the arc joining the $i$ and $(i+1)$th insertion.
These expressions naturally
agree with those obtained from the worldline formalism of Polyakov, Strassler etc.
\polbk\strass\schb . In \ff\ we used these expressions to study the gluing up for
bilinears in the free theory.

A beautiful feature of parametric representations is the
correspondence with electrical networks, originally discovered in
Bjorken's 1958 thesis (see chapter 18 of \BD ). If we identify the
external (as well as internal) momenta with currents flowing in
the network corresponding to the Feynman diagram, then the
Schwinger parameters play the role of resistances. In fact, the
Gaussian exponent, before carrying out the momentum integrals, has
the interpretation as the power dissipated in the original circuit
($\sum_r I_r^2R_r$). The process of carrying out the integrals
over internal or loop momenta is then equivalent to the standard
procedure of elimination of internal currents using Kirchoff's
laws. The resulting Gaussian in the external momenta, given in
\gauss , then has the interpretation as the power dissipated  in
the equivalent circuit after elimination of the internal loops.
This gives us a nice source of intuition for the process by which
loops can get glued into trees. In \ff\ we exploited this to
understand the gluing of the two, three and four point functions
of bilinears into trees.

As we will now see, the correlators $\la \prod_{i=1}^n \Tr\Phi^{J_i}(x_i)\ra_{conn}$
will exhibit the gluing much more completely. In particular,
considering these general correlators will allow us to see all the string moduli,
something which was not possible with bilinears alone, for reasons that
will become clear as we proceed.

\subsec{Gluing into Skeleton Graphs}

In the free theory, the correlators \ncorr\ are given by a
sum over all possible connected Wick contractions. Let us start by
considering the leading large $N$ contribution. They are given by
planar diagrams such as those shown in fig. 1.\foot{In the figure,
the maximal number of connections compatible with planarity have
been drawn. Adding a line between two vertices that are not
already directly connected will destroy planarity.} We have as
many legs coming out of the $i$th vertex as there are free fields
inserted there, namely $J_i$. These planar diagrams are more
easily visualised as spherical diagrams -- drawn on a sphere. How
do we organise the sum over all the different possible
contributions?

Firstly, for a planar graph with
a given connectivity (i.e. the set of pairs of 
vertices $(ij)$ which are linked by at least one 
contraction compatible with planarity),
there can be a multiplicity $m_r$ in the number of lines between each pair.
In fact, one can convince oneself that a planar graph, with $n$ vertices, 
that is maximally connected,
has $3(n-2)$ inequivalent connections. Where the $r$th connection comprises
of $m_r$ lines. $m_r$ is
only constrained by the fact that there must be a total of $J_i$ lines entering
the $i$th vertex. These $n$ constraints imply that there are $2(n-3)$ undetermined
numbers amongst the $m_r$. 
For $n>3$ there is thus a lot of mutliplicity for a
given connectivity.
Secondly, the above multiplicity was for
a fixed connectivity but
it is clear that there are several inequivalent ways (for $n>4$)
to connect the vertices themselves, consistent with planarity.

What we will show in this section is that the first set of
contributions -- from the multiplicity of lines -- can all be
bunched up in a natural way. For a given connectivity, at first it
might seem that the parametric representation \scal\ implies very
different contributions for graphs with differing $m_r$'s,
since we would have to introduce Schwinger parameters for each
internal line. However, we will argue that each of these
contributions can be written in terms of a reduced set of
Schwinger parameters $\t^{eff}_r$ where $r$ runs over the edges in
the corresponding skeleton graph. This skeleton graph is what we
term the graph that captures the connectivity of a given Feynman
diagram.\foot{{\it Caveat}: In order that the skeleton graph faithfully 
capture the colour flow of the original diagram, we will only glue 
together adjacent strips of the underlying double line graph. Lines,
between the same pair of vertices, but
which cannot be deformed into each other without crossing a line between 
a different pair, will {\it not} be glued together. Hence the skeleton
graph could have {\it several} edges between a given pair of vertices. 
Each such edge comes with its own multiplicity. The simplest illustration
of such instances is in the four point function where one can 
have two contractions along
one of the diagonals  (on opposite sides of the sphere, so to say), 
while having none on the other diagonal. Note that such
a graph also has six edges just like the tetrahedron, where all pairs of vertices
are singly connected.} In other words, we replace all the $m_r$ lines in 
a connection by a single edge.
In fig.1 we have illustrated this for our
example. In other
words, all contributions of a given connectivity are expressed in
terms of an integral over parameters defined on the corresponding
skeleton graph. The dependence on the multiplicities $m_r$ is
captured by the integrand in a fairly simple manner.
The nett result is that the skeleton graph and its moduli captures
all the contributions of a given connectivity.

We will argue for this result from the explicit form of the parametrisation
in \scal . However experts might not need much convincing about the truth of this
assertion. (They are welcome to skip the technicalities and go to Eq.(2.9)).
In the Schwinger parameter representation \scal , that we are working with, 
the result
can be understood from the electrical network intuition. In this language, all
we are doing is to replace all the parallel resistors joining vertices $(ij)$
(subject to the caveat in footnote 8) by an effective resistance 
given by the usual
expression for multiple parallel resistors. In that sense, we are partially gluing
up the original Feynman diagram by bunching up various internal lines.

Let's now see how this is reflected in the actual expressions. We
start with a $n$-vertex free field diagram whose connectivity is
specified by a skeleton graph having multiplicity $m_r$ for the
$r$th edge. We will label the Schwinger parameters
for the internal lines by $\t_{r \m_r}$ where $r$ indexes the
edges of the skeleton graph ($r=1\ldots 3(n-2)$) and $\m_r$ their
multiplicity ($\m_r=1\ldots m_r$).

Our first claim relates the term $\D(\t)$ of the original graph to
that of the skeleton graph
\eqn\effmeas{\D(\t)={\prod_{r,\m_r}\t_{r
\m_r}\over\prod_{r}\t^{eff}_r} \tilde{\D}(\t^{eff}).}
Here the effective Schwinger parameter is given by the formula for
parallel resistors
\eqn\effsch{{1\over \t^{eff}_r}=\sum_{\m_r=1}^{m_r}{1\over
\t_{r\m_r}}.}
While $\tilde{\D}(\t^{eff})$ is given by the same expression as
\meas\ but now the sum over 1-trees is that of the skeleton graph
with the effective parameters $\t^{eff}_r$ for the edges. Our
claim follows from the definition in \meas . We are instructed to
take the product of the parameters on the cut lines of the
original graph. In the $r$th bunch, we are forced to cut either
$(m_r-1)$ or all $m_r$ of the lines to get a 1-tree. Any fewer cut
lines would leave a loop. If we were to cut all of them, then we
would get a factor of $\prod_{\m_r}\t_{r \m_r}$ for that bunch and
in the skeleton graph we would have thus cut the corresponding
edge. If we were to cut $m_r-1$ of them, then we would get a
factor $(\prod_{\m_r}\t_{r \m_r})/\t^{eff}_r$ corresponding to all
the possible ways of cutting $(m_r-1)$ lines in that bunch. In the
skeleton graph we would be leaving the $r$th edge uncut. Now it is
clear, from the relative factor of $\t^{eff}_r$ between the two
cases, that on summing over all possible 1-trees of the original
graph we will end up with a sum over 1-trees of the skeleton
graph, obtaining the relation in \effmeas .

The next claim is that the Gaussian exponent in \gauss\ of the
original graph can be expressed entirely in terms of the skeleton
graph with parameters $\t^{eff}_r$. 
\eqn\effgauss{P(\t,k)=\tilde{P}(\t^{eff}, k),} 
where $\tilde{P}(\t^{eff}, k)$ is
given by the same expression as in \gauss , but now for the
skeleton graph with its effective Schwinger parameters for its
edges. This follows from similar considerations as above. The term
in \gauss\ involving the sum over 2-trees is related by a factor
of $(\prod_{r,\m_r}\t_{r \m_r})/\prod_{r}\t^{eff}_r$ to the
corresponding sum over 2-trees of the skeleton graph with
$\t^{eff}_r$ for its edges. The reasoning is completely analogous
to that of the previous para. Putting this together with the
relation \effmeas\ between the factors of $\D$ and $\tilde{\D}$,
we see that the factor of $(\prod_{r,\m_r}\t_{r
\m_r})/\prod_{r}\t^{eff}_r$ cancels out and we are left with the
relation stated in \effgauss .

Putting both these results together, we have for a diagram of
fixed multiplicity and connectivity, the contribution
\eqn\contr{\int_0^{\infty}\prod_{r,\m_r}{d\t_{r\m_r}\over\t_{r
\m_r}^{d\over 2}} {\prod_{r}{\t^{eff}_r}^{d\over 2} \over
\tilde{\D}(\t^{eff})^{d\over 2}}\exp{\{-\tilde{P}(\t^{eff},
k)\}}.} The final step is to convert this into an integral over
the $\t^{eff}_r$. Since the non-trivial dependence in the
integrand is all on the $\t^{eff}_r$, the dependence on the
$\t_{r\m_r}$ can be factored out by a change of variables. The
details are worked out in Appendix A. The end result
is that the contribution \contr\ to the $n$ point function \ncorr\
from a graph with fixed connectivity and multiplicity can be
written as
\eqn\fincont{\int_0^{\infty}\prod_{r=1}^{3(n-2)}
{C^{(m_r)}d\t_r\over \t_r^{(m_r-1)({d\over 2}-1)}}
{1\over\D(\t)^{d\over 2}} \exp{\{-P(\t , k)\}}.}
Here $C^{(m_r)}$ is a constant, independent of the $\t$'s but
depending on $m_r$, obtained from the change of variables in
Appendix A. It is explicitly given by
\eqn\jac{C^{(m_r)}=\int_0^1\prod_{\m_r=1}^{m_r}
dy_{\m_r}y_{\m_r}^{{d\over 2}-2}\d(1-\sum_{\m_r}y_{\m_r}).} Note
that in the interesting case of $d=4$, $C^{(m_r)}={1\over
(m_r-1)!}$.

We have also dropped the superscript on the $\t$'s as well as the
tildes. Hopefully this will not create any confusion, since from
now on only the effective Schwinger parameters will play a role.
Furthermore, all quantities such as $\D(\t)$ and $P(\t, k)$ will
refer to the skeleton graph.

Therefore we can write the total planar contribution to the
momentum space version of \ncorr\ in the form
\eqn\totcont{\eqalign{G^{\{J_i\}}(k_1, k_2,\ldots
k_n)=&\sum_{skel. graphs}
\sum_{\{m_r\}=1}^{\infty}\prod_{i=1}^n\d_{\Sigma m_{r(i)},J_i}
\prod_r C^{(m_r)} \cr \times &\int_0^{\infty}\prod_{r=1}^{3(n-2)}
{d\t_r\over \t_r^{(m_r-1)({d\over 2}-1)}} {1\over\D(\t)^{d\over
2}} \exp{\{-P(\t, k)\}}.}} The sum is over various inequivalent
planar skeleton graphs with $n$ vertices. The sum over
multiplicities is constrained by the fact that the net number of
legs at the $i$th vertex  is $J_i$. ($r(i)$ labels an edge which
has the $i$th vertex as one of its endpoints.)

Thus we see that the planar $n$-point correlator can be written as
an integral over the space of planar skeleton graphs. By this we
mean that \totcont\ includes both an integral over the length of
the edges (as parametrised by the $\t$'s) of a given skeleton
graph, as well as a sum over the different ways of joining the
$n$-vertices. In the next section we will show that this space is
the same as that of the moduli space of a sphere with $n$ holes.
We will also look at various generalisations.

\newsec{From Skeleton Graphs to String Diagrams}

\subsec{Skeleton Graphs and the Cell Decomposition of Moduli Space}

To see the string theory emerge from the field theory, we need to
have the space of string diagrams arise from the field theory
Feynman graphs. By making a correspondence of the above space of
planar skeleton graphs with ${\cal M}_{0,n}$ (and more generally
${\cal M}_{g,n}$), we will accomplish precisely that.

The correspondence is made by observing firstly that the space of
$n$-vertex planar skeleton graphs, which we have been considering, is
nothing other than the space of all triangulations of the sphere
with $n$-vertices. When we say triangulations we mean that the
maximum number of edges, consistent with planarity, namely
$3(n-2)$, arise when all the faces of the skeleton graph are
triangles. If one of the faces of the discretised sphere were not
a triangle, we could always add at least one extra edge without
destroying planarity. In other words, the region of parameter
space where quadrilaterals  and other polygons appear in the faces
are codimension one or higher in the parameter space. More
precisely, quadrilaterals etc. arise only when one or more of the
$\t_r$ go to $\infty$. That is because the corresponding edges are
effectively removed since the resistance in those edges is going
to $\infty$.  In sum, associated uniquely, to every discretisation
of the sphere with $n$ vertices is  a planar skeleton graph
arising from a Feynman diagram and vice versa.

Now, to each such discretisation of the sphere with $n$ vertices
we can uniquely associate a dual graph in the standard manner.\foot{We 
would like to thank S. Wadia for a helpful remark about
the relation between graph duality and open-closed string
duality.} Namely, to each edge of the original graph we associate
a dual edge which intersects the original one transversally. We
will also associate a length $\s_r\equiv {1\over \t_r}$
("conductance") to this edge. The length of individual dual edges
can then vary in an unconstrained manner from $0$ to $\infty$ as
we vary $\t_r$. In this way, every face of the original graph
gives rise to a vertex for the dual and vice versa. The dual graph
is thus constrained to have $n$ faces. And corresponding to the
triangular faces are now trivalent vertices. But the topology
remains that of a sphere. Therefore, as we sum over inequivalent
skeleton diagrams, we carry out a sum over the space of dual
graphs.\foot{The field theory correlators in \ncorr\ are 
usually taken to be those of normal ordered operators. In such a case 
there are no self contraction diagrams. In the correspondence to dual
graphs, self contractions lead to tadpole subgraphs. Presumably there
exists a redefinition on the $AdS$ side which corresponds to the normal
ordering prescription on the field theory side. This would then take care of 
the tadpole diagram contributions in the cell decomposition of the 
moduli space.}
That is, over all discretisations of the sphere with $n$
faces formed from graphs with cubic vertices. As mentioned
earlier, the lengths $\s_r$ of the edges of the dual graph
vary from $0$ to $\infty$.

This can immediately be recognised as the picture of string
interactions in Witten's Open String Field Theory \witcube . Open
string field theory generates string diagrams described by strips
of fixed width but varying lengths $\s$, meeting at cubic vertices. In
fact, as shown first by \gmw\ and argued later in full generality
by \zwie , such diagrams of arbitrary genus, with some number of
boundaries as well as punctures, precisely generate a single cover
of the corresponding moduli space of Riemann surfaces with
boundaries and punctures. This ``cell (or simplicial) decomposition" 
of the moduli
space was also worked out independently 
by mathematicians \penn .\foot{We would like to 
thank P. Windey and S. Govindarajan for
suggesting early on, a possible connection between the approach of
\ff\ and the work of Penner \penn .
For a nice introduction to Penner's work, see the recent article \sunil .} 
Thus the sum over inequivalent 
skeleton graphs is the sum over different cells in this decomposition 
of the moduli space.

An important aspect of the cell decomposition is the way different
components in this decomposition of the moduli space connect to
each other across boundaries of these cells.\foot{We would like to
thank A. Sen for helpful discussions on this point.} It can be
verified that the mapping to dual graphs preserves this behaviour.
For example, in the case of the four point function, consider an
original skeleton graph in the shape of a tetrahedron with all six
legs of non-zero length. One can go to a codimension one boundary
of the cell where the length $\s$ of one of the dual edges goes to
zero. This corresponds in the original graph to removing an edge and
getting a quadrilateral 
face.
From this boundary one can move to a component in which the
edge opposite to it (i.e. having no vertex in common with it)
develops a second strand but now traversing the opposite side of
the sphere (see footnote 8). Mapping this onto the dual graphs
one exactly gets the matching up of the different  
codimension one components
of
the cell decomposition of ${\cal M}_{0,4}$. Similarly one can go
to codimension two boundaries of this codimension one cell and see
that they also patch together smoothly. In all cases graph duality
faithfully implements the required behaviour.

Thus, we can conclude that the space of planar skeleton graphs
with $n$ vertices is isomorphic to the moduli space ${\cal
M}_{0,n}$ of a sphere with $n$ boundaries (faces). Seen in this
light, the lengths of the $3(n-2)=n+2(n-3)$ edges of the original
graph (and thus of the dual graph) correspond to the number of
moduli of a sphere with $n$ holes. In conformal field theory
language one associates $n$ of these to the radii of the holes and
$2(n-3)$ to the positions of centres. As described in the
introduction and illustrated in fig.2, the appearance of ${\cal
M}_{0,n}$ is what one might expect from open-closed duality.

Here we should make a remark regarding the $J_i$ that appear in
\ncorr . For an $n$ point function, unless the $J_i$ are greater
than a minimum value (set by $n$) the Feynman graph will not have
all the possible contractions. In other words, the corresponding
skeleton graph will not have the maximal number of edges i.e.
$3(n-2)$. One concludes that such an amplitude gets its
contribution from a lower dimensional component of the cells of
${\cal M}_{0,n}$. In particular, we see that the bilinear
operators don't get contributions from the whole of the string
moduli space. For example, the Feynman graph for the four point
function of bilinears has only four edges. Thus it gets its support only from
a codimension two slice of ${\cal M}_{0,4}$.

\subsec{Generalisations}

We should also remark that the argument of the present section
only relied on the existence of skeleton graphs. The procedure by
which the skeleton graphs themselves arose from the underlying
field theory diagrams also appears to generalise to operators
other than the scalars $\Tr\P^J$. The parametric
representation for diagrams involving more general operators only
differ in having additional (momentum and spin dependent) polynomial
prefactors multiplying the same Gaussian factor $P(\t,k)$ of \gauss . General
expressions for these prefactors are given in \lam\lama\ (see, in particular, 
equations (11)-(15) of \lam). When one takes into 
account the fermions, gauge fields and global quantum numbers that 
a theory like ${\cal N}=4$ Yang-Mills posseses, the explicit expressions 
for general operators become quite cumbersome. However, 
an examination of the general parametric form in \lam\ reveals that the 
gluing arguments of Sec. 2 generalise for such diagrams as well.
In fact, this is to be expected from the correspondence with
electrical networks, which holds very generally. The only difference is 
that the information about the spins and field content of operators 
now modifies the first term in \fincont . 
Therefore, it appears that 
general (planar) $n$-point correlators in free field
theory can also be expressed as integrals over ${\cal M}_{0,n}$.

Again, the restriction to planar graphs was also not very
essential to the whole argument. The gluing into skeleton graphs
makes no reference to the underlying genus. It is important, however, 
that the gluing be carried out compatible with the colour flow as outlined 
in footnote 8.
Graphs corresponding
to higher genus Feynman diagrams are then glued up into skeleton graphs
which are discretisations of Riemann surfaces with more handles.
Similarly, the mapping to dual graphs gives rise to string
diagrams that cover the moduli space ${\cal M}_{g,n}$. As we
remarked earlier, the cell decomposition of \gmw\zwie\penn\ holds
for any genus $g$ Riemann surfaces with $n$ holes.

Finally, we should also remark that once we have completely understood the free  
field theory (at least in the case of ${\cal N}=4$ Yang-Mills) as a string theory,
we can hope to generalise our approach to the interacting theory. At least, 
order by order in perturbation theory in the Yang-Mills coupling,  
the effect of the coupling is through insertions of additional operators in 
correlators. Since the parametric representation is 
applicable to the corresponding
Feynman diagrams of the interacting theory, we can write it again in terms of an 
integral over a string moduli space but now with additional holes for the coupling
constant insertions. 
It should then be possible to view these additional insertions
as changing, for instance, the radius of the $AdS$. 
In this way, this procedure may be useful in
tackling the $AdS/CFT$ conjecture beyond the free limit as well. 

It is satisfying that our arguments are not too tied up either
with the specifics of the correlators or that of the planar limit (or even too
much with the free limit).
It suggests a universality that behoves the phenomenon of field
theory/string theory duality. Also the fact that the spacetime
dimension does not play a crucial role at this level is also not such a
bad thing. It is a feature which we expect will mostly affect the
integrand over moduli space. The integrand holds the key to the
real dynamics of the string theory which we see emerging from the
field theory. In the next two sections we will make some preliminary
stabs at the integrand, leaving a detailed study for later.

\newsec{Remarks on the Integrand}

The primary result of this paper (specialising to the concrete example 
of scalars) is that we can rewrite field theory correlators, schematically, as 
\eqn\scalstr{G^{\{J_i\}}(k_1, k_2,\ldots k_n)=
\int_{{\cal M}_{g,n}}[d\s]\r^{\{J_i\}}
(\s)\exp{(-\sum_{i,j=1}^n g_{ij}(\s)k_i\cdot k_j)}.}
Here we are denoting the coordinates on the moduli space ${\cal M}_{g,n}$
collectively by $\s$. Recall that $\s_i={1\over \t_i}$ were the natural 
coordinates in the cell decomposition of ${\cal M}_{g,n}$. $\r^{\{J_i\}}(\s)$
is the momentum independent prefactor which captures the dependence on the 
$J_i$. Whereas $g_{ij}(\s)$ in the exponent is independent of the $J_i$. We 
can write down $\r^{\{J_i\}}(\s)$ and $g_{ij}(\s)$ in each 
cell of the moduli space from the expressions at the end of Sec. 2. 

Thus, for instance in the interesting case of $d=4$, 
in a particular cell labelled by a given skeleton graph, 
we can rewrite the contribution in \totcont\ in several equivalent ways
\eqn\cellcont{\eqalign{G^{\{J_i\}}_{cell}=&\sum_{\{m_r\}=1}^{\infty}
\prod_{i=1}^n\d_{\Sigma m_{r(i)},J_i}
\int_0^{\infty}\prod_r
{d\s_r\s_r^{m_r-1}\over(m_r-1)!} 
{1\over\hat{\D}(\s)^2} \exp{\{-\hat{P}(\s, k)\}}\cr
=&\sum_{\{m_r\}=1}^{\infty}\int_0^{\infty}\prod_r
{d\s_r\s_r^{m_r-1}\over(m_r-1)!}\int_0^{2\pi}\prod_{i=1}^n d\theta_i
e^{i\theta_i(\Sigma m_{r(i)}-J_i)}{1\over\hat{\D}(\s)^2} \exp{\{-\hat{P}(\s, k)\}}\cr
=&\int_0^{\infty}\prod_r
d\s_r
{\exp{\{-\hat{P}(\s, k)\}}\over\hat{\D}(\s)^2}\int_0^{2\pi}\prod_{i=1}^n d\theta_i
\exp{(\sum_{r(ij)}\s_{r(ij)}e^{i(\theta_i+\theta_j)})}
e^{-i\sum_{i=1}^n\theta_i(J_i-N_i)}.}}
To obtain the first line we have changed variables in \totcont\ to 
$\s_r={1\over \t_r}$ and rexpressed both $\D(\t)$ and $P(\t,k)$ in terms 
of the $\s$'s. In the process, we have defined 
\eqn\newmeas{\hat{\D}(\s)\equiv\sum_{T_1}(\prod\s)
=(\prod_r\s_r)\D(\t=1/\s).}
and 
\eqn\newgauss{\hat{P}(\s,k)\equiv{1\over\hat{\D}(\s)}
\sum_{T_2}(\prod\s)(\sum k)^2=P(\t=1/\s,k).}
The sum, as before, is over the 1-trees and 2-trees of the skeleton graph
but the product in both these 
definitions is over the lines that are {\it not} cut.

In the second line of \cellcont\ we introduced a lagrange multiplier for
the constraints on the multiplicities. This enables us to carry out the sum 
over multiplicities in an unconstrained way and obtain the third line.
Here $r(ij)$ is an edge that joins vertices $i$ and $j$; and $N_i$ is the number 
of legs joining at the $i$th vertex of the skeleton graph. In this last line 
the cell contribution is clearly in the form \scalstr . From \cellcont\
it is also not difficult to verify that the integrand is continuous 
across boundaries of the cells (where at least one of the $\s\rightarrow 0$).
This is crucial if one wants to interpret the integrand as that of 
a string theory. 

We also notice that the schematic form \scalstr\ is 
similar in structure to the expressions for string 
amplitudes that one is familiar 
with, such as in flat space. Namely, a prefactor contains
the information about the 
masses/dimensions (and more generally spins). While the Gaussian factor is 
independent of these details and captures the (worldsheet) correlators of the 
vertex operators $e^{ik\cdot X(\xi)}$. We will see in the next section that one
can plausibly argue that this is also the structure one would expect from stringy
correlation functions in $AdS$.  

An important feature of string amplitudes is its factorisability in different
channels. This holds at the level of the integrand on moduli space, since it is 
a consequence of the worldsheet OPE and its associativity. Now, in 
the $AdS/CFT$ conjecture, the factorisability of $AdS$ amplitudes is reflected 
in the spacetime OPE relations for the corresponding correlation functions. 
Associativity of the OPE means 
that we can factorise it in different channels yielding the same answer.  
We believe that the above Schwinger 
parametric representation should reflect the spacetime OPE of the 
field theory and hence translate into a factorisability of the integrand 
in the closed string channel. 
It should be very much possible to make this statement precise. 

Ultimately, one wants to also demonstrate that the integrand is specifically 
that of an appropriate string theory on $AdS$. We expect that the details 
of the string theory will depend on the matter content of the field theory.
However, any string theory that is dual to a free (and thus conformal) 
gauge theory should have a background which contains at least an $AdS$ part.  

In \ff\ we pointed out that the appearance of $AdS_{d+1}$ from a free theory 
in $d$ dimensions could naturally take place in the Schwinger
representation that we have been employing. 
Essentially, propagators in $AdS_{d+1}$ can be parametrically 
expressed in terms of $d$ dimensional proper time propagators 
for free fields. 
We used this fact, together with the 
geometric gluing into trees, to argue that the 
two and three point functions of bilinears can be rewritten as 
tree amplitudes on $AdS$. This was accomplished by a simple change of variables 
on the Schwinger parameter space. 

In Sec. 5 we generalise this to the planar three 
point function of $\Tr\P^{J_i}$. What will be clear from the details of 
that calculation is that (as in \ff ) the three Schwinger 
moduli transmute into parameters for the 
propagators on the external legs of the $AdS$ amplitude. Integrating over these
parameters is integrating over the size of the holes of ${\cal M}_{0,3}$. It 
effectively gives rise to punctures in that one gets bulk to boundary $AdS$ 
propagators as a result. This is in line with the intuition, mentioned 
in the introduction, of holes closing 
up as one integrates over their size modulus. 
Together with the appearance of the string moduli space, this gives confidence 
that we are indeed seeing the $AdS$ emerge from the field theory.

From the form of correlators in $AdS$ (discussed in Sec.5)  
we expect this to continue to happen for the $n$ point function. Namely, one can 
isolate $n$ size moduli out of the $6g+3(n-2)$ moduli. And these will simply 
parametrise the $n$ external legs of the corresponding $AdS$ amplitude. The rest
of the integral over the moduli space would then give a closed string $n$-point
amplitude on $AdS$.

\newsec{The Three Point Function and $AdS$ Correlators}

\subsec{From Delta to Star}

We will consider the $n=3$ case of \ncorr\ (in the planar sector) 
\eqn\nthree{G^{\{J_i\}}(k_1, k_2, k_3)=\la \Tr\Phi^{J_1}(k_1)
\Tr\Phi^{J_2}(k_2)\Tr\Phi^{J_3}(k_3)\ra_{conn}.} 
The analysis is a 
generalisation of that in \ff\ but will be done
in a somewhat different way to make things clearer.\foot{See \fmmr\lmrs\dfs\hsw 
etc. for studies of 
three point functions of such scalars (chiral primary operators in the 
${\cal N}=4$ theory)
in the context of $AdS/CFT$.}

The first thing to note, in this case, is that the number of legs 
$m_r$ in the $r$th edge are determined
completely by the $J_i$. 
In fact, we have three equations (from the three vertices) 
\eqn\multp{m_{12}+m_{13}=J_1}
and cyclic permutations of it. Here we are labelling the edges $r$ by the pair of 
vertices they connect. The equations \multp\ determine the $m_{ij}$ to be 
\eqn\multsol{m_{12}={1\over 2}\sum_{i=1}^3J_k-J_3}
and cyclic permutations. Thus there is a unique graph contributing to 
\nthree\ with a fixed number of 
legs between each pair of vertices. We do not have to carry out 
any sum over multiplicities.

Now, by the arguments of Sec. 2, this graph can be glued up into a skeleton graph,
which is just a triangle in this case. 
And the expression for the amplitude in terms of
the effective Schwinger parameters is given by \fincont . 
(Since the skeleton graph 
is unique, upto reflection, this is the same as \totcont .) Actually, as in 
\cellcont , we will work with the natural 
conductance variables $\s_r={1\over \t_r}$
and rewrite \fincont , using \newmeas ,\newgauss\  as
\eqn\thrcont{G^{\{J_i\}}(k_1, k_2, k_3)=\int_0^{\infty}\prod_{r=1}^3
d\s_r\s_r^{(m_r-1)({d\over 2}-1)+{d\over 2}-2}{1\over\hat{\D}(\s)^{d\over 2}} 
\exp{\{-\hat{P}(\s, k)\}}.}
Here we have relabelled the edges so that $\s_1\equiv \s_{23}$ etc. and dropped 
the overall factors of $C^{(m_r)}$.  
Also using the expressions \newmeas\ and \newgauss\ we have 
\eqn\thrdel{\hat{\D}(\s)=\s_1\s_2+\s_2\s_3+\s_3\s_1 }
and 
\eqn\thrp{\hat{P}(\s, k)={1\over\hat{\D}(\s)}[\s_1k_1^2+\s_2k_2^2+\s_3k_3^2].}

We will now re-express this in terms of new moduli, more appropriate for the tree 
\eqn\varch{{1\over \r_i}={\s_i \over \hat{\D}(\s)} 
\Rightarrow \s_i={\r_1\r_2\r_3\over 
(\sum_k\r_k )\r_i}.}
This change of variables is motivated by the star-delta transformation 
of electrical 
networks. Namely, if $\s_i$ are the conductances of a delta or triangle network, 
such as the one we have, then $\r_i$ are the conductances of the equivalent three 
pronged tree or star network (see \guil\ for example). 
It can be checked that the jacobian for this transformation is given by
\eqn\sjac{det({\p\s_i\over \p\r_j})={\r_1\r_2\r_3\over 
(\sum_k\r_k )^3}.}
We also see that 
\eqn\delrho{\hat{\D}(\s)={\r_1\r_2\r_3\over (\sum_k\r_k )}; 
~~~~~~~~~ \hat{P}(\s, k)
=\sum_{i=1}^3{k_i^2\over \r_i}.}

We can now rewrite the integral in \thrcont , after gathering together 
various terms, 
\eqn\rhocont{\eqalign{G^{\{J_i\}}(k_1, k_2, k_3)=&\int_0^{\infty}\prod_{i=1}^3
d\r_i\r_i^{(\Sigma_k m_k-m_i)({d\over 2}-1)-{d\over 2}-1}{1\over 
(\sum_k\r_k )^{\Sigma_k m_k({d\over 2}-1)-{d\over 2}}}
\times e^{-[\sum_{i=1}^3{k_i^2\over \r_i}]}\cr
=&\int_0^{\infty}\prod_{i=1}^3
d\r_i\r_i^{\D_i-{d \over 2}-1}{1\over 
(\sum_k\r_k )^{\Sigma_k {\D_k\over 2} -{d \over 2}}}
\times e^{-[\sum_{i=1}^3{k_i^2\over \r_i}]}.}}
In the second line we have used \multp ,\multsol  
as well as the the fact that the 
operators $\Tr\P^{J_i}$ have canonical dimensions $\D_i=J_i({d \over 2}-1)$ in 
the free theory. 

This last line is close to what one might expect from a string theory on 
$AdS$, as we will shortly see. In any case, it is a short step now to write
\rhocont\ in terms of the expected bulk-to-boundary propagators in $AdS$.   
\eqn\btob{G^{\{J_i\}}(k_1, k_2, k_3)=\int_0^{\infty}{dt\over t^{{d\over 2}+1}}
\int_0^{\infty}\prod_{i=1}^3
d\r_i\r_i^{\D_i-{d \over 2}-1}t^{\D_i\over 2}e^{-t\r_i}e^{-{k_i^2\over \r_i}}.}
Here we have used the identity 
\eqn\useid{{1\over a^s}={1\over \G(s)}\int_0^{\infty}dt t^{s-1}e^{-at}.}
to rewrite the denominator term in \rhocont .

Either in this form or after a Fourier transform to position space we can 
recognise this to be the product of three bulk to boundary propagators in 
$AdS_{d+1}$ for the appropriate scalar fields. 
Thus, for instance in position space, (taking into account the overall 
momentum conserving delta function), we can write \btob\ as 
\eqn\posbtob{\eqalign{G^{\{J_i\}}(x_1, x_2, x_3)=&
\int_0^{\infty}{dt\over t^{{d\over 2}+1}}\int d^dz
\int_0^{\infty}\prod_{i=1}^3d\r_i\r_i^{\D_i-1}t^{\D_i\over 2}
e^{-\r_i(t+(x_i-z)^2)}\cr
=&\int_0^{\infty}{dt\over t^{{d\over 2}+1}}\int d^dz
\prod_{i=1}^3 K_{\D_i}(x_i,z;t),}}
where 
\eqn\btobdef{K_{\D}(x,z;t)={t^{\D\over 2}\over [t+(x-z)^2]^{\D}}}
is the usual position space bulk to boundary propagator for a scalar field 
corresponding to an operator of dimension $\D$. 
The only slight difference
is that we have parametrised the $AdS$ radial coordinate by $z_0^2=t$ as in \ff .

What we have thus seen here is that the integral over the moduli space 
${\cal M}_{0,3}$, 
which the parametric representation of field theory provided us,
is really an integral over $AdS$. The original Schwinger parameters $\s_i$
can be traded for the $\r_i$ which parametrise the propagators for the external 
legs of the $AdS$ correlator. Integrating over the $\r_i$, which correspond to
the size of the holes, propagates the $AdS$ scalar field all the way from
infinity (the boundary). This very much corresponds to the picture in the 
introduction of the holes being replaced by punctures. We will see below how this 
is likely to be more general than to the three point function. 

\subsec{Vertex Operators in $AdS$} 

We can also understand how \thrcont\ or equivalently \rhocont\
could arise from a vertex 
operator calculation in $AdS$. Though we don't have a good handle yet on the 
string theory, we can guess that the $n$-point correlators are 
given in terms of vertex operator computations in the worldsheet theory for $AdS$.
Thus for scalars we would guess, following \pol\tse
\eqn\vert{\eqalign{G^{\{J_i\}}(x_1 \ldots x_n)=&\la\prod_{i=1}^n 
K_{\D_i}(x_i,X(\xi_i);t(\xi_i))\ra_{WS}\cr
=& \la\prod_{i=1}^n{t(\xi_i)^{\D_i\over 2}\over [t(\xi_i)+(x_i-X(\xi_i))^2]^{\D_i}}\ra_{WS}.}}
Here $X(\xi), t(\xi)$ denote worldsheet fields for the $AdS$ target space. The 
averaging, as the subscript indicates, is over the worldsheet action for 
these and other fields (including ghosts). An integral over the moduli space
of the Riemann surface with $n$ punctures is also implicit.
We can write \vert\ in the parametric form
\eqn\adspar{G^{\{J_i\}}(x_1 \ldots x_n)=
\int_0^{\infty}\prod_{i=1}^n d\r_i\r_i^{\D_i-1}
\la t(\xi_i)^{\D_i\over 2}e^{-t(\xi_i)\r_i-\r_i(x_i-X(\xi_i))^2}\ra_{WS}.}
To make a connection with the field theory expressions we go to momentum space 
where \adspar\ becomes
\eqn\adsmom{G^{\{J_i\}}(k_1 \ldots k_n)=
\int_0^{\infty}\prod_{i=1}^n d\r_i\r_i^{\D_i-{d\over 2}-1}e^{-{k_i^2\over \r_i}}
\la t(\xi_i)^{\D_i\over 2}e^{-t(\xi_i)\r_i}e^{i k_i\cdot X(\xi_i)}\ra_{WS}.}

We believe \adsmom\ is the right starting point for a comparison of the (scalar) 
$n$-point function in $AdS$ with the field theory expressions \totcont\ etc.. 
But we can already see 
over here many of the features that we expect. There are $n$ parameters $\r_i$
which can be identified with the size moduli of holes as we argued at the end of 
the last subsection. Then there are the usual $(6g+2n-6)$
moduli for the $n$ point function. As in the case of the three point function we 
need to find the appropriate change of variables to go from these 
parameters to the $(6g+3n-6)$
$\s_i$ of the field theory. But it is clear that \adsmom\ fits in with the 
general schematic form of \scalstr .

In the particular case of the three point function that we studied above, 
since \adsmom\ should be independent of the $\xi_i ~(i=1\ldots 3)$ from conformal
invariance, it is plausible that only the zero mode of the fields $t(\xi), X(\xi)$
effectively contribute in the worldsheet path integral 
(after including the contribution of appropriate ghost insertions). 
The zero mode for $t$ gives the 
corresponding integral in \btob\ and that for $X$ just gives the overall momentum 
conserving delta function. Thus it is not surprising from this point of view that 
we could relate the field theory three point function to the point particle
amplitude \posbtob\ -- only the zero modes contribute. 
It also suggests that we will really see the stringy structure in the
four and higher point functions. 

Going by the arguments presented in this paper, 
the field theory expressions such as \totcont\ or \cellcont\ are just 
\adsmom\ written
in different variables. So we can use this to
turn things around and write down the $AdS$ 
correlators from the field theory (certainly in the case of 
${\cal N}=4$ Super Yang-Mills theory). We would then have 
reconstructed the string theory on $AdS$ 
via all its correlators. 

Anyhow, the task now is 
obviously to make various of these surmises precise
and in the process learn about the worldsheet theory for $AdS$.
In some sense we are in a situation very similar to that in the early 
days of dual theory
when people reconstructed the string picture from the form of the 
Veneziano-Koba-Nielsen and Virasoro-Shapiro amplitudes. 

\medskip
\centerline{\bf Acknowledgements}

I would like to acknowledge the useful comments and questions from the participants 
of the National Workshop on String Theory at I.I.T. Kanpur (Dec. 2003), 
where a preliminary version of this work was presented. I would like to specially thank 
Ashoke Sen for a number of very helpful discussions. Finally, I must express my 
gratitude to the generosity of my fellow Indians in supporting the enterprise of 
theoretical physics. 

\appendix{A}{A Change of Variables}

Here we will see how to effect the change of variables of integration
from the $\t_{r\m_r}$ in \contr\ to the effective Schwinger parameters $\t_r$ 
in \fincont .
Firstly, the relation between $\t_r$ and $\t_{r\m_r}$ is given by \effsch .
This can be implemented by inserting into the integral \contr\ the identity
\eqn\ident{\int_0^{\infty}{d\t_r\over \t_r^2}\d({1\over \t_r}-\sum_{\m_r=1}^{m_r}
{1\over\t_{r\m_r}})=1.}
The nontrivial dependence on $\t_{r\m_r}$ in \contr\ comes from the term in 
the first bracket. So using the above identity we can write such a contribution as
\eqn\modint{\int_0^{\infty}{d\t_r\over \t_r^2}\int_0^{\infty}\prod_{\m_r}
{d\t_{r\m_r}\over\t_{r\m_r}^{d\over 2}}\d({1\over \t_r}-\sum_{\m_r=1}^{m_r}
{1\over\t_{r\m_r}}).}
Now define $x_{r\m_r}={\t_{r\m_r}\over \t_r}$ and change variables from $\t_{r\m_r}$
to $x_{r\m_r}$. Then \modint\ reads as
\eqn\modinta{\eqalign{&\int_0^{\infty}{d\t_r\over \t_r^{m_r({d\over 2}-1)+2}}
\int_1^{\infty}
\prod_{\m_r}{dx_{r\m_r}\over x_{r\m_r}^{d\over 2}}\d[{1\over \t_r}(1-\sum_{\m_r=1}^{m_r}
{1\over x_{r\m_r}})]\cr =&\int_0^{\infty}{d\t_r\over \t_r^{m_r({d\over 2}-1)+1}}
\int_1^{\infty}
\prod_{\m_r}{dx_{r\m_r}\over x_{r\m_r}^{d\over 2}}\d(1-\sum_{\m_r=1}^{m_r}
{1\over x_{r\m_r}}).}}
Thus we have factored the integral over $\t_{r\m_r}$ into an integral over $\t_r$ times 
a factor $C^{(m_r)}$ which depends only on $m_r$, where
\eqn\cmr{\eqalign{C^{(m_r)}=&\int_1^{\infty}
\prod_{\m_r}{dx_{r\m_r}\over x_{r\m_r}^{d\over 2}}\d(1-\sum_{\m_r=1}^{m_r}
{1\over x_{r\m_r}})\cr
=&\int_0^1\prod_{\m_r=1}^{m_r}
dy_{r\m_r}y_{r\m_r}^{{d\over 2}-2}\d(1-\sum_{\m_r=1}^{m_r}y_{r\m_r}).}}
In the second line we have made the substitution $y_{r\m_r}={1\over x_{r\m_r}}$.
In this form we can do the integral explicitly for general $d$. 
But the  
case of $d=4$ is particularly simple. The 
delta function over one of the $y_{r\m_r}$
can be carried out and we are left with an integral over 
the $(m_r-1)$ others in 
over the region where their sum is less than one. This 
is just ${1\over(m_r-1)!}$. In general dimensions the answer is
\foot{We would like to thank E. Schreiber for providing us with this 
expression.}
\eqn\cmd{C^{(m_r)}= {\G({d\over 2} - 1)^{m_r} \over\G(m_r({d\over 2} - 1))}.}

\listrefs

\end